\documentclass{ewic-latex}
\usepackage{url}
\usepackage{hyperref}

\begin{document}

\runningheads{Axelsson $\bullet$ Sikau}{Choreographing Trash Cans: On Speculative Futures of Weak Robots in Public Spaces}

\conference{}

\title{Choreographing Trash Cans: On Speculative Futures of Weak Robots in Public Spaces}

\authorone{Minja Axelsson*\\
Designer and HRI Scholar\\
University of Cambridge\\
UK\\
\email{mwa29@cam.ac.uk}}

\authortwo{Lea Luka Sikau*\\
Media Artist and Posthuman Studies Scholar \\
Die Junge Akademie \\
GE\\
\email{leasikau@outlook.de}}

\begin{abstract}

Delivering groceries or cleaning airports, mobile robots exist in public spaces. While these examples showcase robots that execute tasks, this paper explores mobile robots that encourage posthuman collaboration rather than managing environments independently. With feigned fragility, cuteness and incomplete functionalities, the so-called ''weak robots'' invite passersby to engage not only on a utilitarian level, but also through imaginative and emotional responses. After examining the workings of ''weak robots'' by queering notions of function and ability, we introduce two speculative design fiction vignettes that describe choreographies of such robots in future urban spaces---one exploring a utopian weak robot and the other a dystopian weak robot. We introduce these speculations in order to discuss how different values may drive design decisions, and how such decisions may shape and drive different socio-technical futures in which robots and humans share public spaces that incentivise collaboration.

\end{abstract}


\maketitle

\section{Introduction: Framing this Work}

This paper explores the performative interplay between humans and ``weak robots" in public spaces. Michio Okada first conceptualised ``weak robots'', which have limited capabilities themselves, and are framed as objects or ``social others'' which people are invited to assist and take care of. In Okada's work, such robots are used to invite pro-social behaviour from people, such as encouraging them to pick up trash to assist a trash can robot (\cite{okada2022weak}). We conceptualise human–robot interaction (HRI) as a stage where weak robots---designed to be ``cute'' and vulnerable---play the role of incidental actors that subvert the person engaging with them. \cite{caudwell2020cute} argue that cuteness as a design choice for robots can encourage users to trust and form relationships with those robots, which introduces ambivalent power dynamics through the production of intimacy. In fact, cuteness can also be seen as a deceptive or ``dark'' pattern, due to the utilisation of cuteness to prompt affective responses which can be used to collect emotional data, as well as some degree of reduction of user agency (\cite{caudwell2019cute}). The ability and affordances of cute and weak robots to influence user behaviour merits the discussion of their ethicality, which we do in this paper through design fiction.

Unlike traditional HRI research, often confined to laboratory settings, our focus is on spontaneous, real-world interactions that transform everyday environments into sites of performative potential. We argue that the theatricality of these encounters is central to understanding their impact: the presence of a weak and/or cute robot, such as the trash can robot, developed by Okada and the Interaction and Communication Design Lab of the Toyohashi University of Technology, acts as a disruptive interloper that introduces an observer's effect and, thus, affects the human interlocutors. 

First, we examine the concept of weak robots through the lens of performativity theory as well as concepts of machine (dys)function. Then, we contextualise notions of "weakness" and "cuteness" through disability studies and queer theory. We then present our design fiction vignettes of the future, and identify what values are embedded and expressed in each design, what design decisions and guidelines were followed, and what outcomes are driven by the design. In the context of these vignettes, we explore both the utopian and dystopian utilisation of cuteness and weakness in robot design, in order to provoke thought on the ethicality of these human-robot interactions. Finally, we discuss our vignettes and the weak robot concept through the lens of performativity theory, examining what functions are revealed and concealed by the robotic mask. 

\section{Weak Robots and (Dys)function}

Robots are increasingly present in public spaces as mobile, semi-autonomous agents. These include delivery robots on sidewalks
\footnote{e.g., Starship technologies on UK's streets, \url{https://www.cambridge-news.co.uk/whats-on/cambridge-delivery-robots-stalemate-face-28462078}, Accessed: 24-06-2025} autonomous cleaning machines in transport hubs\footnote{e.g., Ice Clean's Co-Botics at a UK airport \url{https://www.roboticsandautomationmagazine.co.uk/news/cleaning-sanitation/robots-enhance-london-luton-airport-cleaning-regime.html}, Accessed: 24-06-2025}, and trash-collecting robots in parks and stations (\cite{okada2022weak}). While typically described in terms of efficiency and automation, these robots also function as public interfaces of ability. Their presence invites passersby to assess them on specific expectations about action, assistance, and competence.

As Mara Mills and Jonathan Sterne have argued, the history of technological design is deeply entwined with regimes of ability and impairment, often defining "function" against an imagined norm of bodily or cognitive autonomy (\cite{mills2011disability, sterne2012mp3}). Weak robots (\cite{okada2022weak})---machines that simulate helplessness, slowness, or partial dysfunction---subvert these assumptions by strategically embodying incapacity. They invite interaction not through ``mastery'' or speed, but through breakdown of function. Simultaneously, weak robots play with the notion of ``cuteness'' as a strategy to cause affective reactions of the incidental user. These robots are not errors in the system; they are designed to perform a failure of executing what they appear to be made for. Weakness here is used as a ``'strength' that enriches the relationship'' (\cite{ikumi2023creating}). 

The design choice of weakness destabilises the binary between function and dysfunction, moving toward a relational and contingent understanding of ability. As Aimi Hamraie notes in their work on access and design, normate space is structured around invisible assumptions of independence and legibility (\cite{hamraie2017building}). Public robots, particularly weak ones, test these assumptions. Their ``failures'' make space for new modes of attention, timing, and care. A robot that pauses, stalls, or requires assistance alters not only pedestrian flows but the temporality of public behaviour. In this way, it enacts what Alison Kafer might call ``crip time'': a reorientation of normative pace and relationality (\cite{kafer2013feminist}).

The performativity of robots in public thus becomes a site for negotiating ability and access. For instance, when a delivery robot got stuck in snow, a passerby empathised with the ``poor little mite'', and ``rescued'' it, so that it could continue its delivery\footnote{The BBC: ``Cambridge delivery robot grateful for snow rescue'' \url{https://www.bbc.co.uk/news/uk-england-cambridgeshire-63997868}, Accessed: 24-06-2025}. The passerby remarked the robot ``thanking'' them, and how it ``looked lost''. This is an example of a robot inviting anthropomorphisation through the design choices of weakness and cuteness, and through that inviting human action and intervention. In fact, \cite{pelikan2024encountering} found that for robots to enter public spaces, they need to be ``granted passage'' through the active interaction work of community inhabitants. Understanding this dynamic invites critical engagement with the aesthetics and politics of robot design, particularly with regards to embodied (dys)functionalities in shared spaces.

\section{Queering Ability: Disability as Method in Robotic Speculation}

We propose two speculative futures which explore weak robots through the lens of dystopia and utopia. Our speculative design fiction vignettes explore worldmaking shaped through values embedded in robot design. To make sense of these divergent trajectories, we draw on disability studies and queer theory to reframe robotic functioning systems as relational, contingent, and co-constructed. Rather than treating limitations as failures, we apply disability as method (see, e.g., \cite{kafer2013feminist, hamraie2017building}) to view non-functionality in weak robots as a design intervention that reshapes normative expectations of autonomy and efficiency.

Robots that stall or require interpretation and intervention unsettle dominant imaginaries of automation and progress. Their performances unfold in ``crip time'' (\cite{kafer2013feminist})---a temporality shaped by interdependence, interruption, and attunement. These robots do not simulate disability; rather, as Mills argues in her genealogy of assistive technology (\cite{mills2011disability}), they make visible how all technical systems emerge through thresholds of ability. In these vignettes, we explore the role of the robot, the human interactor, and the bystanders as a serendipitous audience who elicit performativity of the human interactor.

In our dystopian vignette, the weak robot enforces surveillance and oppressive normativity under the guise of fragility. In contrast, the utopian vignette mobilises weakness to enable care, autonomy of the human individual, and non-intrusive cooperation with the goal of pro-socially shaping an enjoyable shared public space. This contrast foregrounds how robotic ability is not fixed but negotiated through aesthetic and infrastructural choices. Recent queer HRI research reinforces this view, calling for robotic agents that reflect multiplicity, resist simplification, and support identity beyond normate assumptions (see e.g., \cite{winkle2021flexibility}, \cite{seaborn2022self}, \cite{axelsson2025queer}). 

By treating disability and queerness as methodological lenses rather than user categories, we shift the question from how robots can support difference to how design itself constructs or forecloses certain futures. Weak robots, precisely because of their indeterminacy, become speculative tools to probe what kinds of relations are possible and what kinds of public spaces we are building. In the following section, we utilise these robots as speculative tools through design fiction.

\section{Utopian and Dystopian Design Fiction Vignettes}

We use a framework of identifying: 1) the values embedded and expressed by each vignette (inspired by Value Sensitive Design (\cite{friedman2013value}), in which values are centred in the design process); 2) the design decisions made to reach this future design of a robot; and 3) the outcomes driven by those design decisions. These are made explicit to probe how different values, decisions driven by those values, and outcomes elicited by those decisions materialise in speculative human-robot interactions. We do this also to elicit discussion on how such values, decisions and outcomes may be challenged in the future, to drive different and more ethical futures in which robots share and shape public spaces with humans.

We meet our protagonist Laura (she/her), a 10-year-old commuting alone through a metro station. Each vignette examines the roles of the robot, the human, and the bystanders through alternating perspectives, with special attention to sensory design and embodied aesthetics.

\subsection{A Dystopian Choreography}

\textbf{Design fiction scenario:}

\textit{Human side:}  
Laura walks through the metro station. While searching her pocket for her phone, a tissue falls. Moments later, she hears a sharp, chirping tone. The sound combines childlike laughter with mechanical urgency, attracting immediate attention.

\textit{Robot side:}  
The trash can robot is designed in pastel pink with an exaggeratedly rounded form. Its face-like interface blinks with cartoonish eyes, and its LED ring glows red as it detects the trash. A surveillance camera embedded within its face zooms in and silently begins recognition and data matching.

\textit{Human side:}  
Laura hesitates. The robot begins sobbing audibly in a digitally modulated tone, simulating distress. Its screen flashes a heart emoji that warps into a crying face. The robot tilts forward, spotlighting the trash with an accusatory beam.

\textit{Bystanders:}  
Several commuters pause and turn. The robot loops an animation of a falling tear and trash, accompanied by a jingle that mimics a classroom chime. Some observers frown and whisper.

\textit{Human side:}  
Embarrassed, Laura picks up the tissue and places it into the bin. The robot emits a melodic tone and displays a winking emoji. She bends down to pat it.

\textit{Robot side:}  
The robot records the interaction and transmits Laura’s behavioural data. The apparent emotional exchange is one-sided and instrumental.

\textit{Bystanders:}  
A local shopkeeper in the crowd later recalls Laura’s face, reevaluating her family through the lens of the public reprimand.

\textbf{Values embedded and expressed:}  
Normative enforcement, behavioural control, emotional manipulation via interface design

\textbf{Design decisions and guidelines:}  
Surveillance via facial recognition, use of exaggerated cuteness, misleading emotional affect

\textbf{Outcomes driven by design:}  
Social shaming, emotional investment in deceptive agents, obfuscation of power asymmetry through aesthetic choices


\subsection{A Utopian Choreography}

\textbf{Design fiction scenario:}

\textit{Human side:}  
Laura walks through a metro station. A tissue falls from her coat pocket. She hears a soft tone that resembles wind chimes, diffuse and non-intrusive.

\textit{Robot side:}  
The robot is shaped like a garden planter, textured with recycled material in muted grey tones. It lacks cartoon eyes, and instead moves with subtle mechanical tilts. A glowing green ring at its base pulses slowly as it identifies the object.

\textit{Human side:}  
Laura, familiar with this robot’s gestures from school, understands its intention. The robot circles the trash, emitting a slow glissando tone, framed as invitation rather than directive.

\textit{Bystanders:}  
The interaction is unobtrusive. One passerby nods warmly at Laura. Most others remain unaware of the exchange.

\textit{Human side:}  
Laura deposits the tissue. The robot responds with a celebratory jingle composed of tuned percussion. It spins gently and dims its light as it departs.

\textit{Robot side:}  
The robot does not collect or store personal data. Its gestures remain understated, maintaining a non-relational affective stance.

\textit{Bystanders:}  
The moment dissolves into the background flow of the station. The robot blends into the temporality of the crowd, rather than intervening in it.

\textbf{Values embedded and expressed:}  
Mutual respect, non-invasiveness, ambient sociability, respecting user autonomy, respecting diversity through non-personalisation, privacy of data and from public reprimand

\textbf{Design decisions and guidelines:}  
Transparent operation, minimal affective projection, soft tactility in audio-visual output

\textbf{Outcomes driven by design:}  
Public trust, incidental cooperation, decoupling cuteness from manipulation


\section{Discussion: Weak Robots Queering Future Urbanism}

In both speculative vignettes, the robot’s presence prompts a suspension of disbelief among bystanders. It transforms incidental interactions into ethically and aesthetically charged encounters, a plateau interrogating the nature of performative observation itself.

Drawing on Judith Butler’s theory of performativity (\cite{butler1993bodies}), we consider how weak robots do not simply act within public space but rather enact themselves through relational context. Their meaning and status as helpers, intruders, or social ``others'' emerge through interaction. The robot becomes legible through choreographies of gestures, signals, and behaviours. These performances are not just communicative but constitutive: they create the social roles the robot inhabits.

This aligns with Karen Barad’s concept of intra-action, where agency is not located in individual entities but arises through relational entanglement (\cite{barad2007meeting}). The weak robot's functionality exists in collaboration with its environment: its audience and cultural framing. 
Observers do not merely project meaning onto the robot; they are drawn into a mutual performance, temporarily accepting its limitations and interpreting its intentions, and adjusting their behaviour in response. The robot, by failing to fully meet expectations of autonomy or utility, opens up space for this engagement to occur.

This performativity also has infrastructural dimensions. As Jussi Parikka argues, malfunction and non-functionality in media technologies can render visible otherwise invisible systems of control and expectation (\cite{parikka2012medianature}). Weak robots, by design, leverage these moments of slowness or confusion to activate awareness of the surrounding socio-technical choreography. In our speculative vignettes, the systems of control and expectation are made visible through the identification of values embedded and expressed, the design decisions and guidelines followed, and the outcomes driven by the design. When designing future robots, designers should endeavor to similarly make visible the driving motivations and interests embedded into the performance in which their robots engage people. This will allow designers to think critically about the ethical and socio-technical impact of the human-robot interaction (\cite{axelsson2025four}) and their design decisions, especially if utilising notions of weakness and cuteness, particularly in robots designed for public spaces.

\section{Presentation Format for the Conference}

This paper will be presented in poster format. We will ask participants to take part in an ideation exercise with the format introduced in this presentation: writing a vignette of a utopian or dystopian public robotics future with the actors of a robot, a person, and bystanders. We will then ask them to identify what values are embedded and expressed in their scenario, what design decisions were made and what design guidelines followed, and identify the outcomes driven by design decisions. This method will enable us to discuss with attendees their opinions on what makes public-space robots utopian and dystopian.

\section{Acknowledgements}
* denotes equal contribution by M.A. and L.L.S. M. A. acknowledges a postdoctoral fellowship grant from the Emil Aaltonen Foundation.


\begin{thebibliography}{9}


\bibitem[Axelsson et al. (2025)]{axelsson2025four}
Axelsson, M., Cheong, J., Nyrup, R., \& Gunes, H. (2025). \textit{Who Owns The Robot?: Four Ethical and Socio-technical Questions about Wellbeing Robots in the Real World through Community Engagement.} In Proceedings of the AAAI/ACM Conference on AI, Ethics, and Society. AAAI/ACM.

\bibitem[Axelsson (2025)]{axelsson2025queer}
Axelsson, M. (2025). \textit{Speculative Design of Equitable Robotics: Queer Fictions and Futures}. In Proceedings of the 2025 British Computer Society’s Special Interest Group in Human Computer Interaction Conference (BCS HCI), Futures track.

\bibitem[Barad (2007)]{barad2007meeting}
Barad, K. (2007). \textit{Meeting the Universe Halfway: Quantum Physics and the Entanglement of Matter and Meaning}. Duke University Press.

\bibitem[Butler (1993)]{butler1993bodies}
Butler, J. (1993). \textit{Bodies That Matter: On the Discursive Limits of "Sex"}. Routledge.

\bibitem[Caudwell and Lacey (2020)]{caudwell2020cute}
Caudwell, C., \& Lacey, C. (2020). \textit{What do home robots want? The ambivalent power of cuteness in robotic relationships}. Convergence, 26(4), 956-968.

\bibitem[Friedman et al. (2013)]{friedman2013value}
Friedman, B., Kahn, P. H., Borning, A., \& Huldtgren, A. (2013). \textit{Value sensitive design and information systems.} Early engagement and new technologies: Opening up the laboratory, 55-95.

\bibitem[Hamraie (2017)]{hamraie2017building}
Hamraie, A. (2017). \textit{Building Access: Universal Design and the Politics of Disability}. University of Minnesota Press.

\bibitem[Ikumi (2023)]{ikumi2023creating}
Ikumi, T. (2023, November). \textit{Creating a prosperous society with `weak robots' that can’t do anything on their own}. Diversity in the Arts Today. \url{https://www.diversity-in-the-arts.jp/en/stories/41805}, Accessed: 24-06-2025.

\bibitem[Kafer (2013)]{kafer2013feminist}
Kafer, A. (2013). \textit{Feminist, Queer, Crip}. Indiana University Press.


\bibitem[Lacey and Caudwell (2019)]{caudwell2019cute}
Lacey, C., \& Caudwell, C. (2019, March). \textit{Cuteness as a ‘dark pattern’ in home robots}. In 2019 14th ACM/IEEE International Conference on Human-Robot Interaction (HRI) (pp. 374-381). IEEE.

\bibitem[Mills (2011)]{mills2011disability}
Mills, M. (2011). \textit{Hearing Aids and the History of Electronics Miniaturization. IEEE Annals of the History of Computing}, 33(2), 24–44.

\bibitem[Okada (2022)]{okada2022weak}
Okada, M. (2022). \textit{Weak robots.} JSAP Review, 2022, 220409.

\bibitem[Parikka (2012)]{parikka2012medianature}
Parikka, J. (2012). \textit{What is Media Archaeology?} Polity Press.

\bibitem[Pelikan et al. (2024)]{pelikan2024encountering}
Pelikan, H. R., Reeves, S., \& Cantarutti, M. N. (2024, March). \textit{Encountering autonomous robots on public streets.} In Proceedings of the 2024 ACM/IEEE International Conference on Human-Robot Interaction (pp. 561-571).

\bibitem[Seaborn (2022)]{seaborn2022self}
Seaborn, K. (2022). \textit{From Identified to Self-Identifying: Social Identity Theory for Socially Embodied Artificial Agents.} In Proceedings of the HRI 2022 Workshop on Robo-Identity, Vol. 2.

\bibitem[Sterne (2012)]{sterne2012mp3}
Sterne, J. (2012). \textit{MP3: The Meaning of a Format}. Duke University Press.

\bibitem[Winkle et al. (2021)]{winkle2021flexibility}
Winkle, K., Jackson, R. B., Bejarano, A., \& Williams, T. (2021). \textit{On the Flexibility of Robot Social Identity Performance: Benefits, Ethical Risks and Open Research Questions.} HRI Workshop on Robo-Identity.





\end{thebibliography}
\end{document}